\documentclass[letterpaper, 10 pt, conference]{ieeeconf}  % Comment this line out if you need a4paper

\IEEEoverridecommandlockouts                              % This command is only needed if 
\usepackage{cite}
\usepackage{amsmath,amssymb,amsfonts}
\usepackage{algorithmic}
\usepackage{graphicx}
\usepackage{textcomp}
\usepackage{mathtools}
\usepackage{xcolor}
\usepackage{booktabs}
\usepackage[caption=false]{subfig}
\usepackage[ruled,linesnumbered]{algorithm2e}

\def\HS#1{\textcolor{black}{{#1}}}
\def\WW#1{\textcolor{black}{{#1}}}
\SetKw{KwBy}{by}
\allowdisplaybreaks

\title{\LARGE \bf
\HS{Multi-Objective Control Co-design Using Graph Based Optimization for Offshore Wind Farm Grid Integration}
}

\author{Himanshu Sharma$^*$, Wei Wang, Bowen Huang, Thiagarajan Ramachandran, and Veronica Adetola% <-this % stops a space
\thanks{This research was supported by the Energy System Co-Design with Multiple Objectives and Power Electronics (E-COMP) Initiative, under the Laboratory  Directed Research and Development (LDRD) Program at Pacific Northwest National Laboratory (PNNL). PNNL is a multi-program national laboratory operated for the U.S. Department of Energy (DOE) by Battelle Memorial Institute under Contract No. DE-AC05-76RL01830.}
\thanks{Email: {\tt\small \{himanshu.sharma, w.wang, bowen.h,
thiagarajan.ramachandran, veronica.adetola\} @pnnl.gov}}\\%
\authorblockA{Pacific Northwest National Laboratory, Richland, WA, USA}
}

\begin{document}

\maketitle
\thispagestyle{empty}
\pagestyle{empty}

%%%%%%%%%%%%%%%%%%%%%%%%%%%%%%%%%%%%%%%%%%%%%%%%%%%%%%%%%%%%%%%%%%%%%%%%%%%%%%%%
\begin{abstract}
% Offshore wind farms (OWF) are increasingly becoming a popular source of renewable energy. They can generate large amounts of electricity with a relatively small environmental footprint but bring complex challenges when integrated into the power-grid. To address these challenges, it is well understood that optimal-sized energy storage can provide potential solutions and help improve the reliability, efficiency, and flexibility of the grid. However, in the case of OWF, limited studies have tried to perform energy storage sizing while simultaneously including the OWF interconnection controls and operations (i.e., control co-design). Further, it is often identified that the planning requires addressing multiple objectives. Therefore, in the present work, we develop a control co-design optimization formulation such that it optimizes multiple objectives and identifies Pareto optimal solutions. We envision the complexity of the system therefore propose the use of the graph-based optimization framework, which allows us to decompose the optimization problem for large power systems. In our use case, the IEEE-9 bus system is treated as an onshore AC-grid with two OWFs connected to it via a multi-terminal DC grid. The developed methodology successfully identifies the Pareto front during the control co-design optimization and allows the decision maker to select the best compromise solution for multiple objectives.
Offshore wind farms have emerged as a popular renewable energy source that can generate substantial electric power with a low environmental impact. However, integrating these farms into the grid poses significant complexities. To address these issues, optimal-sized energy storage can provide potential solutions and help improve the reliability, efficiency, and flexibility of the grid. Nevertheless, limited studies have attempted to perform energy storage sizing while including design and operations (i.e., control co-design) for offshore wind farms. As a result, the present work develops a control co-design optimization formulation to optimize multiple objectives and identify Pareto optimal solutions. The graph-based optimization framework is proposed to address the complexity of the system, allowing the optimization problem to be decomposed for large power systems. The IEEE-9 bus system is treated as an onshore AC grid with two offshore wind farms connected via a multi-terminal DC grid for our use case. The developed methodology successfully identifies the Pareto front during the control co-design optimization, enabling decision-makers to select the best compromise solution for multiple objectives.
\end{abstract}

%%%%%%%%%%%%%%%%%%%%%%%%%%%%%%%%%%%%%%%%%%%%%%%%%%%%%%%%%%%%%%%%%%%%%%%%%%%%%%%%
\section{INTRODUCTION}
Offshore wind farms (OWF) are gaining increasing attention worldwide for sustainable energy development. In 2022, 8,385 MW of new projects were commissioned for offshore wind energy globally\cite{OffshoreWindReport23}. In the U.S., offshore wind energy production \WW{capacity} potential reached 52,687 MW in 2023, showing a growth of 15\% \cite{OffshoreWindReport23} \cite{OffshoreWindReport23}. With the increasing power extraction from OWFs, it becomes important to develop capabilities to transmit this power efficiently. Most modern OWFs are developed with multi-terminal DC (MTDC) grid with modular multilevel converters (MMC) due to various advantages over high-voltage alternating current (HVAC) lines. A detailed review of MMC-MTDC grid can be seen in \cite{ansari2020mmc}.

Large renewable energy source integration %such as offshore wind 
brings challenges %such as stability and inflexibility 
to the AC-grid operators. A comprehensive discussion is presented in \cite{simao2017challenge}. One of the well-acknowledged approaches to ensure power system stability is battery energy storage systems (BESS) and their operations and control \cite{zhao2023grid}.

% High-voltage direct current (HVDC) transmission is becoming more popular than high-voltage alternating current (HVAC) for long-distance offshore wind integration since AC-cable have high per-km capacitance and higher losses in comparison to DC cables. In addition, HVDC allows to decouple the OWF from the onshore grid, which avoids any voltage instabilities that can arise due to wind power volatility \cite{wei2011voltage}. The HVDC line transmission is achieved through multiple point-point connections with the AC grid, resulting to a multi-terminal setup (MTDC). Most of the modern OWFs are now developed with MTDC grids with modular multilevel converters (MMC)
% technology. A detailed review of MMC MTDC grid can be seen in \cite{ansari2020mmc}. \\ 
% The HVDC connections are based on converter technologies such as line-commutated converters (LCC's) , voltage-source converters (VSCs) and modular multilevel converters (MMC). MMC converters are preferred, as they provide exceptional performance, controllability, scalability and are simple to be applicable for multi-terminal HVDC setup.
% halwany2022optimal,paul2019multi,brekken2010optimal,dui2017two

BESS sizing on AC/DC side is an important decision during OWF interconnection planning. A comprehensive review focusing on determining optimal sizing for wind farm applications can be seen in \cite{Zhao2015ReviewOE}. Recent work by Halwany et al. \cite{halwany2022optimal} developed an approach for doing storage sizing for OWF black start operations with probabilistic approach for onsite energy storage. However, the work did not consider the onshore BESS sizing and interconnection MTDC grid. Santanu et al. \cite{paul2019multi} also proposed a multi-objective approach for battery sizing in OWF considering economic and reliability objectives. The work developed a sequential approach of handling multiple objectives but neglected the battery controls and converter dynamics. Moghaddam et al. \cite{moghaddam2017predictive} considered the BESS sizing problem for an onshore wind farm; however, the authors developed a sequential approach where they first chose BESS size and then proposed a control strategy. To the best of the authors' knowledge, none of the studies has considered simultaneously accounting for the control operations of offshore wind farms while sizing BESS. \\
\indent Conventionally, the design problem is solved first followed by operation control optimizations. However, many studies \cite{allison2014,deshmukh2016,yan2009,liu2020decentralized,bhattacharya2021,nash2021hierarchical,nash2021robust} have shown that such a sequential approach results in sub-optimal system performance. Control co-design (CCD) is a control system design approach that takes into account the interactions between the control system and the underlying design of the physical system.\HS{ A comprehensive review of CCD and handling uncertainties in the formulation presented in \cite{garciasanz2019} and \cite{azad2023overview} respectively}. In this paper, we aim to develop a CCD \HS{approach} that is suitably used for BESS design for the OWF. Specifically, we are interested in developing CCD approach to handle the challenge of large system CCD with tight coupling amongst the sub-systems (i.e MMC's etc.). We aim to develop a generalized framework to pose a co-design optimization problem that can also handle sub-system level coupling constraints during the control and design optimization. Furthermore, many energy systems design requires the system to satisfy multiple objectives (e.g., low operation cost, minimize power loss etc.).
\HS{Few studies on co-design control (CCD) considering multiple objectives have been conducted without identifying the Pareto front \cite{allison2024open,sundarrajan2021towards,allison2014co}. Inspired by recent research on optimizing marine energy kites \cite{naik2023pareto}, we propose a CCD approach that addresses multiple objectives to find Pareto solutions for integrating offshore wind farms into the grid.}
% \HS{Accounting  multiple objectives in CCD have been considered in few studies \cite{allison2024open,sundarrajan2021towards,allison2014co} however,the Pareto front was not identified for the problem. A recent study for optimizing underwater kite to harness Marine hydrokinetic energy present an approach for identifying a Pareto optimal front \cite{naik2023pareto}. We take motivation from the work in \cite{naik2023pareto} to propose a CCD formulations that account for multiple objectives to identify a Pareto solution for multi-objective problem for offshore wind farm grid integration.}
Therefore, in this paper, we aim to address the aforementioned research objective by developing the CCD approach using the graph-based optimization framework, which can allow scalability of optimization in the case of large systems and allow formulating a sub-system level co-design problem. Further, we also develop a gradient-based approach to handle multiple objectives for Pareto set identification.

% Summarize which section means what...!
The rest of the paper is organized as follows. In Section-\ref{sec:Meth}, we discuss the proposed methodology of using graph-based optimization and the gradient based multi-objective optimization approach for Pareto set identification. In Section-\ref{sec:OWF}, we describe the OWF use case for doing CCD and developed optimization formulation. In Section-\ref{sec:results}, we discuss the results from the approach and present our conclusions and future work in Section-\ref{sec:conc}.

\section{Methodology}\label{sec:Meth}
We describe the details on the proposed methodology developed for doing CCD OWF interconnected with AC-grid with BESS. 

\subsection{Graph-Based Optimization}
\begin{figure}
     \centering
     \includegraphics[width=0.8\linewidth,keepaspectratio]{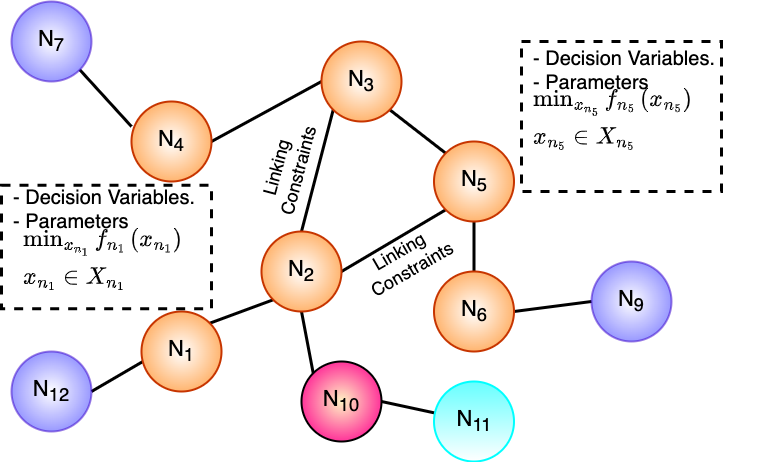}
     \caption{A schematic for the graph-based optimization formulation is shown. Each node defines a sub-system with individual objective functions. }
     \label{fig:GraphMapping}
 \end{figure}

Graph-based modeling abstractions have recently been explored in convex optimization \cite{hallac2017snapvx}, infrastructure networks \cite{jalving2017graph}, supply chain planning problems \cite{berger2021remote}, and simulation of partial differential equations  \cite{abhyankar2018petsc}. These abstractions' structures are directly tied to physical topology of the systems. Recent work by \cite{jalving2017graph,jalving2019graph, jalving2022graph} has shown that the optimization and simulation for complex systems can be represented using the graph-based computational framework. It provides a coherent strategy to capture the modeling elements for a system, which are often common in most engineering applications. Figure \ref{fig:GraphMapping} shows the graph-based representation of optimization problem consisting of a set of nodes and edges. Each node represents an individual sub-system optimization model (with variables, objectives, constraints, and data), and each edge captures connectivity between node models and coupling constraints. 
Once the graph is constructed, it can be communicated to traditional or decomposition optimization solvers (e.g., \textit{Groubi} or \textit{Ipopt}).

%For the presented graph based implementation, a graph $\mathcal{G}(\mathcal{N},\mathcal{E})$ is a collection of nodes $\mathcal{N}$ and edges $\mathcal{E}$. Note that a set of nodes belongs to a specific graph $\mathcal{G}$ and is represented by syntax $\mathcal{N}(\mathcal{G})$ and the node elements are denoted by $n \in \mathcal{N}(\mathcal{G})$. 
In CCD problem for OWF connecting to AC grid through MTDC, we use the system topology of MTDC and AC grids to define the nodes and edges. Each node corresponds to a sub-system level representation. % such as buses, converters, energy storage, branches, etc. In case of current use-case we do not have individual sub-system level objectives and only have multiple system-level objectives for control co-design. Therefore, for each node on the graph, the same objective function were evaluated. 
The details about the CCD formulation and how to convert it to a graph-based model will be discussed in section-\ref{sec:OWF_Form}.

%The system level objectives may have variables not be accessible for a any node to evaluate the objective function. To address this challenge we introduce dummy variables at each node and introduced a linking constraint that ensure those dummy variables at a given node equals the missing variable value.

Next, we describe the details on the proposed gradient-based approach for solving multi-objective optimization.

\subsection{Gradient based Multi-objective optimization}
We propose a gradient-based multi-objective optimization framework inspired by the weighted-sum method and the bi-level optimization algorithm in \cite{gu2022min}. Given $N$ objective functions that can be split into $n$ nodes, the weighted-sum combination of single objective functions $\{f_1,f_2,\ldots,f_n\}$, $\{g_1,\ldots,g_n\}$, $\{h_1,\ldots,h_n\}$ can be written as,
\begin{eqnarray}
    \min &w_1(f_1+f_2+\ldots+f_n)+w_2(g_1+g_2+\ldots+g_n)\nonumber\\
    &+\ldots+w_N(h_1+h_2+\ldots+h_n)\label{eq:moo1}
\end{eqnarray}
where $\{f\},\;\{g\},\;\{h\}$ are groups of individual objective functions into different nodes.
Then the proposed multi-objective optimization framework can be applied to update the weights to find the Pareto frontier, as in the following Algorithm~\ref{alg:moo}.
\begin{algorithm}[htp!]
\caption{Gradient-based Approach for Multi-Objective Optimization}\label{alg:moo}
Set the step sizes $\beta$ for updating $\mathbf{w}$.\;
Solve the graph-based problem with initial weights $\mathbf{w}_0=(w_1,w_2,\ldots,w_N)$ using \textit{Plasmo.jl}.

\For{$k\gets 0$ \KwTo $K$ \KwBy $1$}{
    Update $\mathbf{w}$ with projected gradient descent.
    $\mathbf{w}_{k+1} = \text{proj}_{\Delta N}(\mathbf{w}_k+\beta h_\mathbf{w}^k)$\;
    Solve the graph-based problem with updated weights
    $\mathbf{w}_{k+1}= (w_1,w_2,\ldots,w_N)$ using \textit{Plasmo.jl}
    }
Return $\bar w = w(\tau)$, where $\tau \sim\mathcal{U}(1,\ldots,K)$
\end{algorithm}

As shown in Algorithm~\ref{alg:moo}, the inputs of this algorithm include the initial values of weights $\mathbf{w_0}$, step size $\beta$, and well-defined multi-objective functions, e.g., $(f_1,f_2,\ldots,f_n)$ and $(g_1,g_2,\ldots,g_n)$($N=2$). The weighed-sum of the given objective functions can be written into $n$ nodes graph-based formulation as in (\ref{eq:moo1}). During the iteration from $k=1\sim K$, the weights $\mathbf{w}$ are updated by the step size $\beta$ into the per-objective stochastic gradient estimates $h_\mathbf{w}^k = [\nabla_{\mathbf{w}_k}\mathbf{f},\nabla_{\mathbf{w}_k}\mathbf{g}]$, then the new weights will be its projection to the $N$-simplex defined by $\Delta N \coloneqq \{\mathbf{w}\in \mathbb{R}^N: w_i\geq0, \forall i \in [N], \sum_{i\in[N]}w_i = 1 \}$. For each iteration, the problem will be solved as a single objective graph in \textit{Plasmo.jl} and after $K$ iterations, the output of the algorithm will be uniformly chosen from the generated Pareto frontier weights$(\mathbf{w}_1,\ldots,\mathbf{w}_K)$.

It is imperative to mention that while numerous conventional methods are available to tackle multi-objective optimization problems, %e.g. weighted-sum method, $\epsilon$-constraint method and evolutionary algorithms, 
the reason to choose the proposed gradient-based algorithm is three-fold. Firstly, computational efficiency when solving large-scale graph-based multi-objective optimization problem is still a critical issue, especially when binary variables exist in the problem. The graph-based formulation can be easily integrated in our proposed algorithm for computing its projected gradient without losing its convexity. Secondly, the proposed algorithm supports vector-valued nonlinear objectives and constraints, which can not be directly solved in existing \textit{Plasmo.jl} or \textit{JuMP} multi-objective solvers yet. Finally, this approach is more scalable to graph-based formulation when combined with \textit{Plasmo.jl} to application in large-scale graph network.

\section{Case Study: Offshore Wind Farm Interconnect}\label{sec:OWF}
\begin{figure}[htbp]
\centerline{\includegraphics[width=0.85\linewidth]{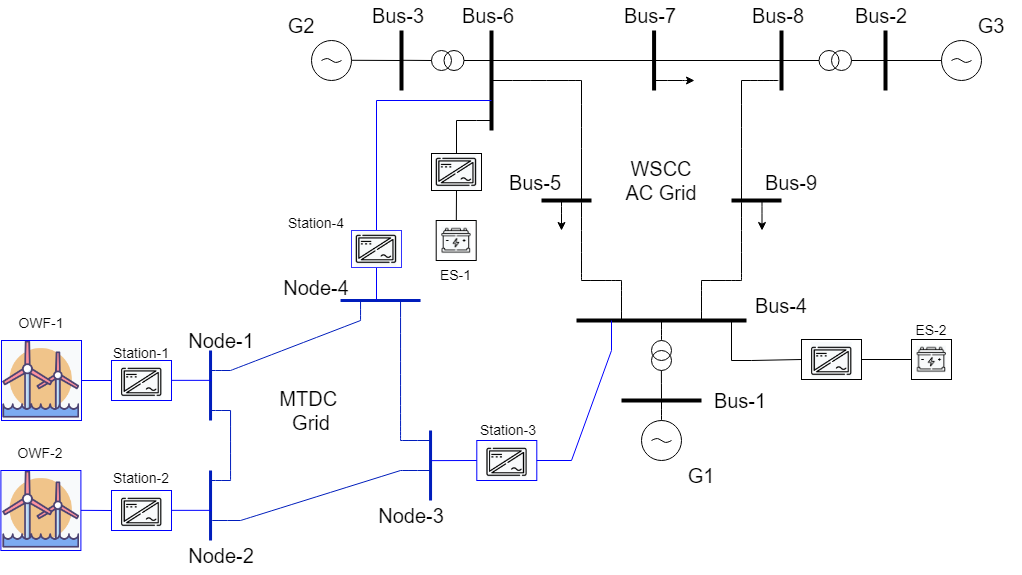}}
\caption{IEEE-9 bus system interconnected with OWFs using MMC-MTDC grid.}
\label{fig_9bus}
\end{figure}

We apply our methodologies on the well-known WSCC 9-bus system. Two OWFs are connected to buses 4 and 6 through a four-terminal MMC-based MTDC network to transmit the wind power\cite{bresesti2007hvdc,liang2011operation}. Two BESS are attached to buses 4 and 6 to help reduce total cost and improve system efficiency. The system structure is shown in Figure \ref{fig_9bus}.

We treat OWFs with converters as active power source and energy storage devices with converters as active power load or sources depending on their charging/discharging operations. MMCs connecting AC and MTDC grids are simply modeled with fixed coefficients\cite{baradar2012power,zhou2019optimal}.% In our work for the offshore AC-DC interconnect system, the controlled power injections from generators, power transmitted from MTDC to AC grid through MMC stations, power flow in both grids, and BESS charge/discharge values are considered as the control (optimization) variables \cite{feng2013new}.

%We formulate our MTDC and AC power flow control co-design problem as a nonconvex optimization problem. However, solving nonconvex problems are quite time-consuming and most nonconvex solvers cannot guarantee the global optimality of their solutions. Therefore, we relax nonconvex MTDC and AC power flow constraints to second order conic program (SOCP) shown in \cite{jabr2006radial,ergun2019optimal}. Translating the whole problem as a SOCP usually gives more accurate solutions than mixed integer linear models. Further, SOCP can be solved by many commercial and open-source optimization solvers and allows to easily apply decomposition algorithms.

The original nonconvex formulation of our MTDC and AC power flow CCD problem is relaxed to a second-order conic program (SOCP)\cite{jabr2006radial,ergun2019optimal}. It usually gives more accurate solutions than linear models and does not have the local optimality issue in solving nonconvex problems. SOCP can either be solved by many optimization solvers or calculated through decomposition algorithms.

\subsection*{Control Co-design Optimization Formulation}\label{sec:OWF_Form}%Control Co-design of OWF Interconnect}
The following two objectives are considered in our model. We can easily extend it to contain additional objetives.
\begin{align}
\min\quad&\sum_{i\in\mathcal S}f^{EI}_i(BS_i)+\sum_{i\in\mathcal B_{AC}}\sum_{t\in\mathcal T}f_i^G(P_{i,t}^G,Q_{i,t}^G)\nonumber\\
&+\sum_{i\in\mathcal S}\sum_{t\in\mathcal T}f_i^{EO}(P_{i,t}^{ch},P_{i,t}^{dis})\label{obj_cost}\\
\min\quad&\sum_{i\in\mathcal C}\sum_{t\in\mathcal T}P_{i,t}^{Loss}+\sum_{i,j\in\mathcal L_{AC}}\sum_{t\in\mathcal T}g_{ij}\left(c_{ii,t}+c_{jj,t}-2c_{ij,t}\right)\nonumber\\
&+\sum_{ij\in\mathcal L_{DC}}\sum_{t\in\mathcal T}g_{DC_{ij}}\left(v_{ii,t}+v_{jj,t}-2v_{ij,t}\right)\label{obj_loss}
\end{align}
\HS{The objective function (eq.~\ref{obj_cost}) is to minimize total cost, including BESS installation $f^{EI}(\cdot)$, regular generator fuel cost, $f^{G}(\cdot ) $ and BESS operation cost $f^{EO}(\cdot)$.} \WW{We use $\mathcal S$, $\mathcal B_{AC}$, and $\mathcal T$ to represent the index sets of batteries, buses in AC grid, and time intervals, respectively}. In the cost functions, $BS_i$ is the size of BESS $i$; $P_{i,t}^{ch}$ and $P_{i,t}^{dis}$ are their charged and discharged power at time $t$, respectively; and $P_{i,t}^G$ and $Q_{i,t}^G$ are active and reactive power output of generator $i$ at time $t$, respectively.

The objective function (eq.~\ref{obj_loss}) is to minimize total power loss, including those in MMCs, AC grid transmission, and DC grid transmission. In this objective function, \WW{$\mathcal C$, $\mathcal L_{AC}$, and $\mathcal L_{DC}$ are index sets of MMCs, branches of AC and MTDC grids, respectively}; variable $P^{Loss}_{i,t}$ represents the power loss in converter $i$ at time $t$; parameters $g_{ij}$ and $g_{DC_{ij}}$ are conductance of AC and DC branch $ij$, respectively; variables $c_{ij,t}$, $v_{ij,t}$, and $s_{ij,t}$ appeared later are used for replacing voltages of AC and DC buses to get SOCP relaxation. Specifically, if we let $E_i$ and $F_i$ be the real and imaginary parts of voltage at AC bus $i$, $V_i$ be the voltage at DC bus $i$, then we have $c_{ij}=E_iE_j+F_iF_j$, $s_{ij}=E_iF_j-E_jF_i$, and $v_{ij}=V_iV_j$. Therefore for each AC or DC branch, they should satisfy the following relationship.
\begin{align}
& c_{ij,t}=c_{ji,t},\ s_{ij,t}=-s_{ji,t}\ \forall ij\in\mathcal L_{AC},\ t\in\mathcal T\label{cs_con}\\
& c_{ij,t}^2+s_{ij,t}^2\leq c_{ii,t}c_{jj,t}\quad\forall ij\in\mathcal L_{AC},\ t\in\mathcal T\label{cs_soc}\\
&v_{ij,t}=v_{ji,t},\ v_{ij,t}^2\leq v_{ii,t}v_{jj,t}\quad\forall ij\in\mathcal L_{DC},\ t\in\mathcal T\label{v_con}
\end{align}

For each bus $i$ at time period $t$, the following active and reactive power balance constraints should be satisfied.
\begin{align}
&P_{i,t}^G-P_{i,t}^D+P_{i,t}^{Conv}-P_{i,t}^{ch}+P_{i,t}^{dis}=\nonumber\\
&\sum_{j\in\mathcal B_{AC}}\left(G_{ij}c_{ij,t}-B_{ij}s_{ij,t}\right)\label{act_bal}\\
&Q_{i,t}^G-Q_{i,t}^D=-\sum_{j\in\mathcal B_{AC}}\left(G_{ij}s_{ij,t}+B_{ij}c_{ij,t}\right)\label{react_bal}
\end{align}
Here parameters $P^D_{i,t}$ and $Q^D_{i,t}$ are active and reactive power demand, $G$ and $B$ are real and imaginary parts of AC grid admittance matrix, variable $P^{Conv}_{i,t}$ is the active power injection from the MMC.

Let $\underline V_i$ and $\overline V_i$ be the lower and upper bounds of the voltage at bus $i$. These limits can be imposed as
\begin{align}
&\underline V_i^2\leq c_{ii,t}\leq \overline V_i^2\quad\forall i\in\mathcal B_{AC},\ t\in\mathcal T\label{ac_vol_lim}
\end{align}

The active and reactive power output of generator $i$ at time $t$ have the following ramping and bound limits, in which the left and right-hand side values are corresponding parameters.
\begin{align}
-Ramp_i^{P-}\leq P_{i,t+1}^G&-P_{i,t}^G\leq Ramp_i^{P+}\label{p_ramp}\\
-Ramp_i^{Q-}\leq Q_{i,t+1}^G&-Q_{i,t}^G\leq Ramp_i^{Q+}\label{q_ramp}\\
P_i^{\min}\leq &P_{i,t}^G\leq P_i^{\max}\label{p_gen_lim}\\
Q_i^{\min}\leq &Q_{i,t}^G\leq Q_i^{\max}\label{q_gen_lim}
\end{align}

Similar to the AC grid, each MTDC bus $i$ at time $t$ has a balance constraint, in which \WW{$\mathcal B_{DC}$ is the index set of buses in MTDC}, $P^{WF}_{i,t}$ is offshore wind power generation, $G_{DC}$ is DC grid admittance matrix, variable $P^{DC}_{i,t}$ is the power injected to MMC.
\begin{align}
P^{WF}_{i,t}-P^{DC}_{i,t}=\sum_{j\in\mathcal B_{DC}}v_{ij,t}G_{DC_{ij}}\label{dc_bal}
\end{align}

Each MMC has three constraints, including power balance, loss estimation, and voltage droop control, which are presented below. The index $i$ on left and right-hand sides represent the AC and DC buses it connects, respectively.
\begin{align}
&P^{Conv}_{i,t}+P^{Loss}_{i,t}=P^{DC}_{i,t}\label{mmc_bal}\\
&P^{Loss}_{i,t}=\beta_i\left|P_{i,t}^{DC}\right|\label{mmc_loss}\\
&(k_iP_{i,t}^{Conv}+d_i)^2\leq v_{ii,t}\label{mmc_ctrl}
\end{align}
Here $\beta$ is the efficiency coefficient, and $k$ and $d$ are droop control parameters.

Finally, we have the constraints for each BESS $i$.
\begin{align}
&SC_{i,t}-SC_{i,t-1}=\eta_i^{ch}P_{i,t}^{ch}-\eta_i^{dis}P_{i,t}^{dis}\label{bat_oper}\\
&0\leq P_{i,t}^{ch}\leq P_i^{ch\_\max}z_{i,t}\label{bat_ch}\\
&0\leq P_{i,t}^{dis}\leq P_i^{dis\_\max}(1-z_{i,t})\label{bat_dis}\\
&0\leq SC_{i,t}\leq BS_i\label{soc_lim}\\
&BS_i^{\min}\leq BS_{i}\leq BS_i^{\max}\label{bs_lim}
\end{align}
Constraint (\ref{bat_oper}) is the operation equation, in which parameters $\eta^{ch}$ and $\eta^{dis}$ are charging and discharging efficiency, respectively. Constraints (\ref{bat_ch}-\ref{bs_lim}) are limits of charging and discharging rates, states of charge, and sizes, respectively.

In this problem, the battery size $BS$ is chosen as our design variable, and the remaining operation-related ones as our control variables, \WW{including charged and discharged power of batteries $P^{ch}$, $P^{dis}$, active and reactive power output of generators $P^G$, $Q^G$, AC and MTDC power flow related variables $c$, $s$, $v$ in constraints (eq.~\ref{cs_con}-\ref{v_con}, eq.~\ref{dc_bal}), and the power through MCC $P^{Conv}$, $P^{Loss}$, $P^{DC}$}.

\begin{figure}[htbp]
\centerline{\includegraphics[width=0.6\linewidth]{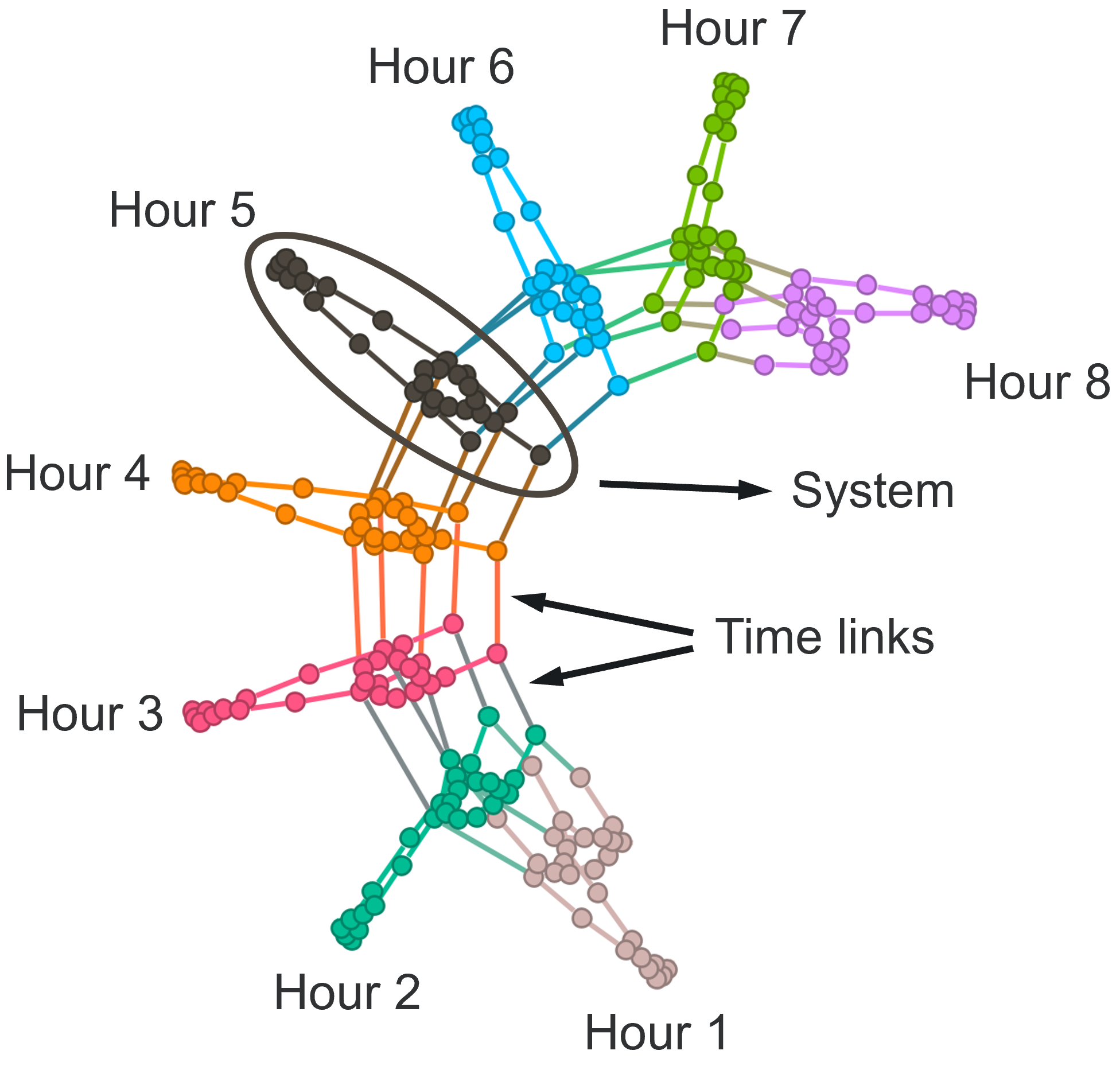}}
\caption{The graph-based co-design optimization problem visualization for IEEE-9 bus-MTDC use case. Each color indicates a time snapshot of the complete system described as a graph node for 8 Hours.}
\label{fig_plasmo}
\end{figure}

To convert this problem into a graph-based model, we define each AC/DC bus, AC/DC branch, converter, and BESS as graph nodes. The linking constraints of the edges and objective terms in each node are derived based the above-described formulation. We consider eight hours as our problem time horizon. The resulting graph is shown in Figure \ref{fig_plasmo} with each color representing an hour. \WW{The structure of the graph shows the nodes with associated decision variables connected in time (sequential hours). While selecting one of the hours on the graph (circled in black) shows the linking constraints among different components, including the same components of the system}.

\section{Results}\label{sec:results}
DC grid branch data is presented in Tables \ref{tab_DCBranch}. The converter droop control and efficiency parameters are $k=0.02$, $d=1$, $\beta=0.03$. The batteries' minimum ($BS^{\min}$) and maximum ($BS^{\max}$) capacities are 20MWh and 120MWh, respectively, and their initial and minimum states of charge at the last hour are set to half of $BS^{\min}$. Batteries' charge and discharge efficiency are $0.8$ and $1.1$, respectively. \WW{Nominal wind farm (WF) outputs are set as 40MW and 50MW, respectively. Nominal load and generator cost functions are taken from the 9-bus system data. In each hour, WF output, nominal load, and fuel cost are multiplied by the factors provided in Table \ref{tab_levels}}. \WW{The problem is formulated in \texttt{Julia} programming with packages \texttt{JuMP} and \texttt{Plasmo}, solved by \texttt{Ipopt} and \texttt{Juniper} on a laptop with i7-12800H CPU and 16GB ram}.

\begin{table}[htbp]
  \centering
  \caption{DC Grid Branch Data}
  \vspace{-0.3cm}
    \begin{tabular}{ccc|ccc}
    \toprule
    From & To & R(p.u.) & From & To & R(p.u.) \\
    \hline
    1     & 2     & 0.0016 & 2     & 3     & 0.0048 \\
    1     & 4     & 0.0048 & 3     & 4     & 0.0042 \\
    \bottomrule
    \end{tabular}%
  \label{tab_DCBranch}%
\end{table}
%\vspace{-0.5cm}
\begin{table}[htbp]
  \centering
  \caption{Load, Wind Farm Output, and Cost Levels}
  \vspace{-0.3cm}
    \begin{tabular}{ccccccccc}
    \toprule
    Hour  & 1     & 2     & 3     & 4     & 5     & 6     & 7     & 8 \\
    \midrule
    Load & 0.9   & 1.1   & 1.25  & 1.4   & 1.55  & 1.3   & 1.15  & 1 \\
    WF Output & 1     & 0.95  & 1.05  & 0.9   & 0.85  & 1     & 1.1   & 0.95 \\
    \WW{Fuel} Cost & 1.1   & 0.9   & 1.3   & 1.5   & 1.8   & 1.6   & 1.4   & 1.4 \\
    \bottomrule
    \end{tabular}%
  \label{tab_levels}%
\end{table}%

\subsection{Single Objective Function}
\begin{figure}[htbp]
\centerline{\includegraphics[width=0.6\linewidth]{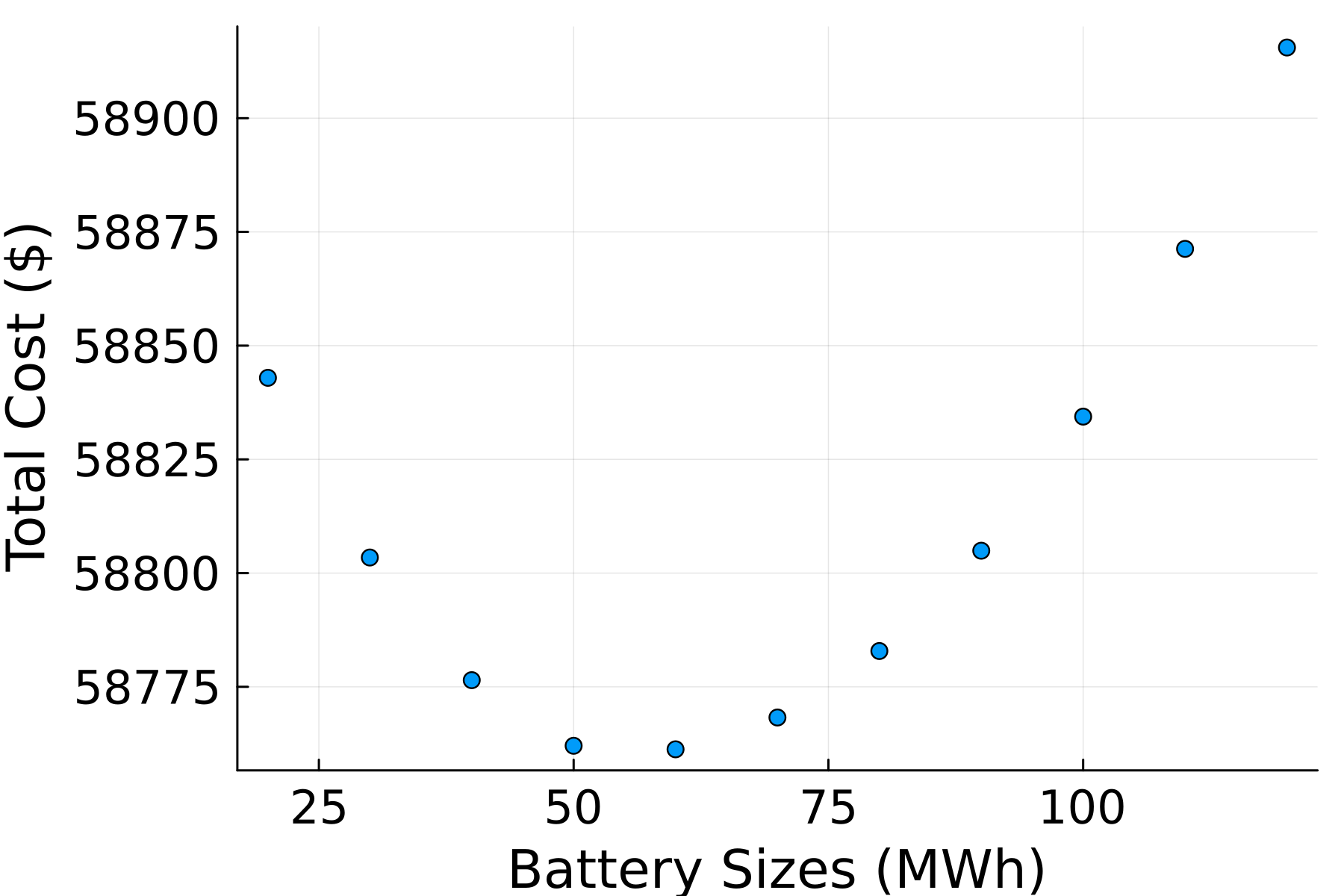}}
\caption{Minimum total system costs with different battery sizes.}
\label{fig_1objBatS}
\end{figure}

We first only consider objective function (eq.~\ref{obj_cost}) for minimizing total cost. The results by \WW{setting each of the two batteries with the same fixed sizes} ranging from 20MWh to 120MWh in 10 MWh increments are shown in Figure \ref{fig_1objBatS}.

% \begin{table*}[htbp]
%   \centering
%   \caption{Minimum Cost with Different Fixed and Variable Battery Sizes}
%   \vspace{-0.3cm}
%     \begin{tabular}{ccccccccccccc}
%     \toprule
%     Battery Size (MWh) & 20    & 30    & 40    & 50    & 60    & 70    & 80    & 90    & 100   & 110   & 120   & Variable \\
%     \midrule
%     Total Cost (\$) & 58843 & 58803 & 58776 & 58762 & 58761 & 58768 & 58783 & 58805 & 58834 & 58871 & 58916 & 58751 \\
%     \bottomrule
%     \end{tabular}%
%   \label{tab_1objBatS}%
% \end{table*}%

By setting battery sizes as decision variables, total cost is minimized \WW{with an optimal solution of 26.5 MWh and 93.2 MWh}. The time for solving this problem is about 45 seconds, while solving all 11 problems with fixed battery sizes takes almost 15 minutes in total.

\begin{figure}[htbp]
\centerline{\includegraphics[width=0.45\linewidth]{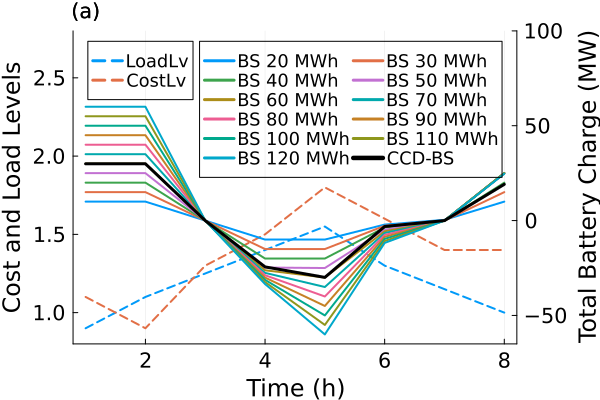}\ \ 
\includegraphics[width=0.45\linewidth]{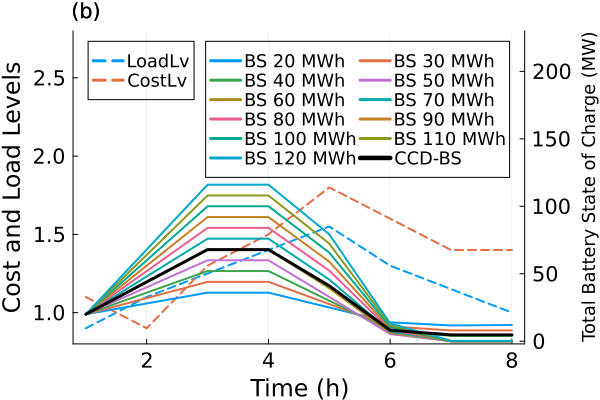}}
\vspace{-0.3cm}
\caption{Different battery sizes and relative cost/load levels in each hour (a) Battery charging/discharging operation (b) Battery state of charge.}
\label{fig_BatCh}
\end{figure}

To better understand how the batteries help reduce total cost, we investigate in detail their hourly operations. \WW{In Figure \ref{fig_BatCh}, the total charging (positive)/discharging (negative) power and state of charge of the batteries are shown by solid lines by the sub-figures, respectively, including results with 11 fixed battery sizes and that of CCD by setting battery sizes as decision variables.} The dashed lines are relative load and generation cost levels in Table \ref{tab_levels}. \WW{The ``BS'' in legend stands for ``Battery Size''}. When load and generation costs are low, like in the first and last few hours, the batteries are charged. These charged power is then used to satisfy demand when load and cost are high. By transferring load from peak to off-peak hours, the total cost is reduced. \HS{By choosing the best battery sizes based on single-objective control co-design}, our model can balance the cost of battery installation, operation, and power generation.

%\begin{figure}[htbp]
%\centerline{\includegraphics[width=0.6\linewidth]{./figs/SoC.png}}
%\caption{Battery state of charge with different battery sizes and relative cost/load levels in each hour.}
%\label{fig_SoC}
%\end{figure}

\begin{figure}[htbp]
\centerline{\includegraphics[width=0.8\linewidth]{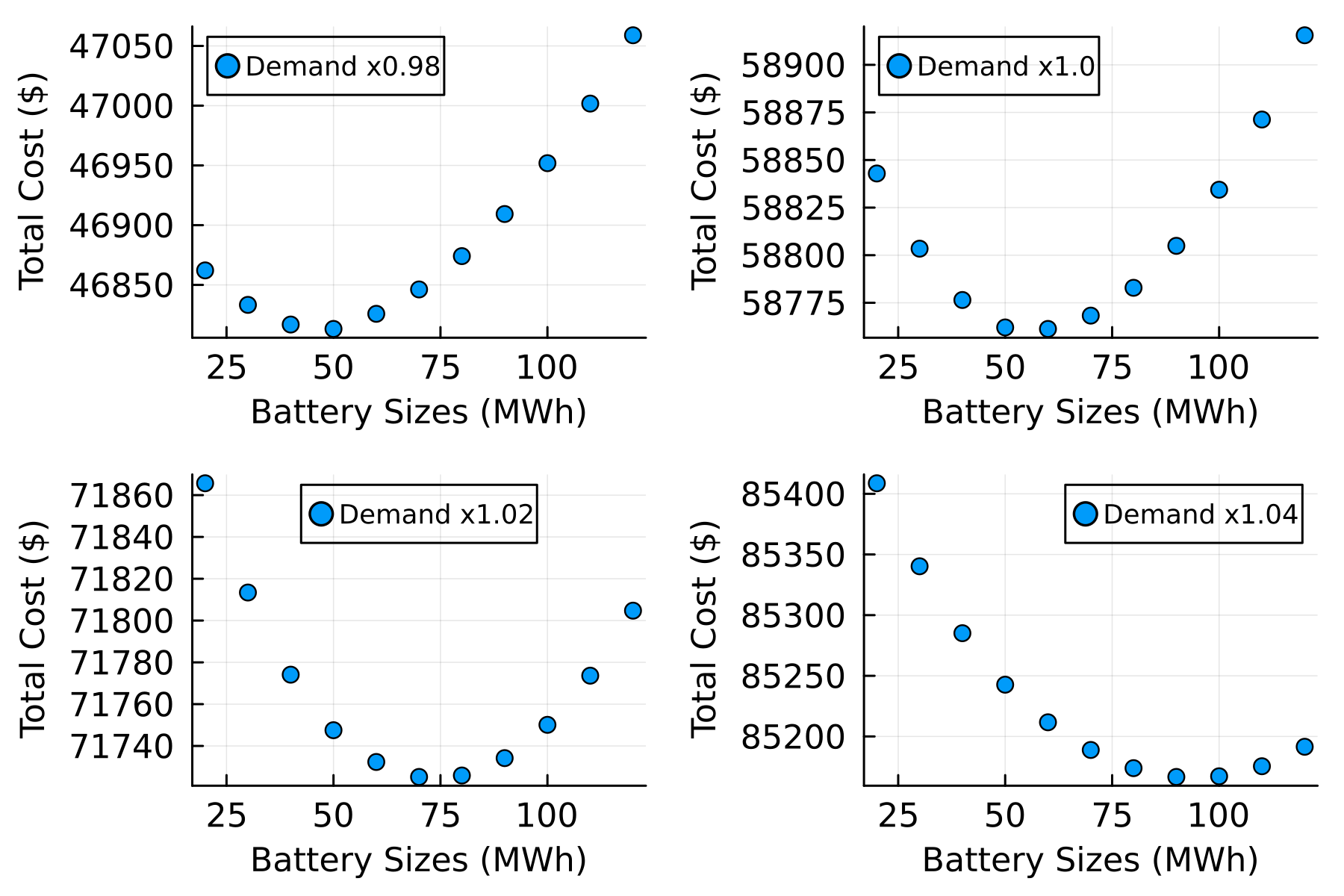}}
\vspace{-0.3cm}
\caption{Total system cost of different nominal demand and battery sizes.}
\label{fig_TotC}
\end{figure}

Since batteries transfer load from peak to off-peak time to reduce cost, it is not difficult to imagine that different load levels may require different battery sizes. We calculate the total cost of the system based on these 11 fixed battery sizes considering demand levels between 98\% and 104\% of the original nominal demand. The results are shown in Figure \ref{fig_TotC}. As the load increases, the best battery sizes to have minimum total cost also increase. On one hand, higher demand may need larger battery capacities for transferring load from peak to off-peak hours. On the other hand, higher generation cost resulting from larger load allows more budgets for battery installation and operation.

\subsection{Multiple Objective Functions}
In this subsection, we minimize both objective functions (eq.~\ref{obj_cost}) and (eq.~\ref{obj_loss}) together. We first simply adopt the linear scalarization approach to find the Pareto front, i.e., we convert the problem as weighted average minimization problem, by defining individual objective weights.

\begin{figure}[htbp]
\centerline{\includegraphics[width=0.5\linewidth]{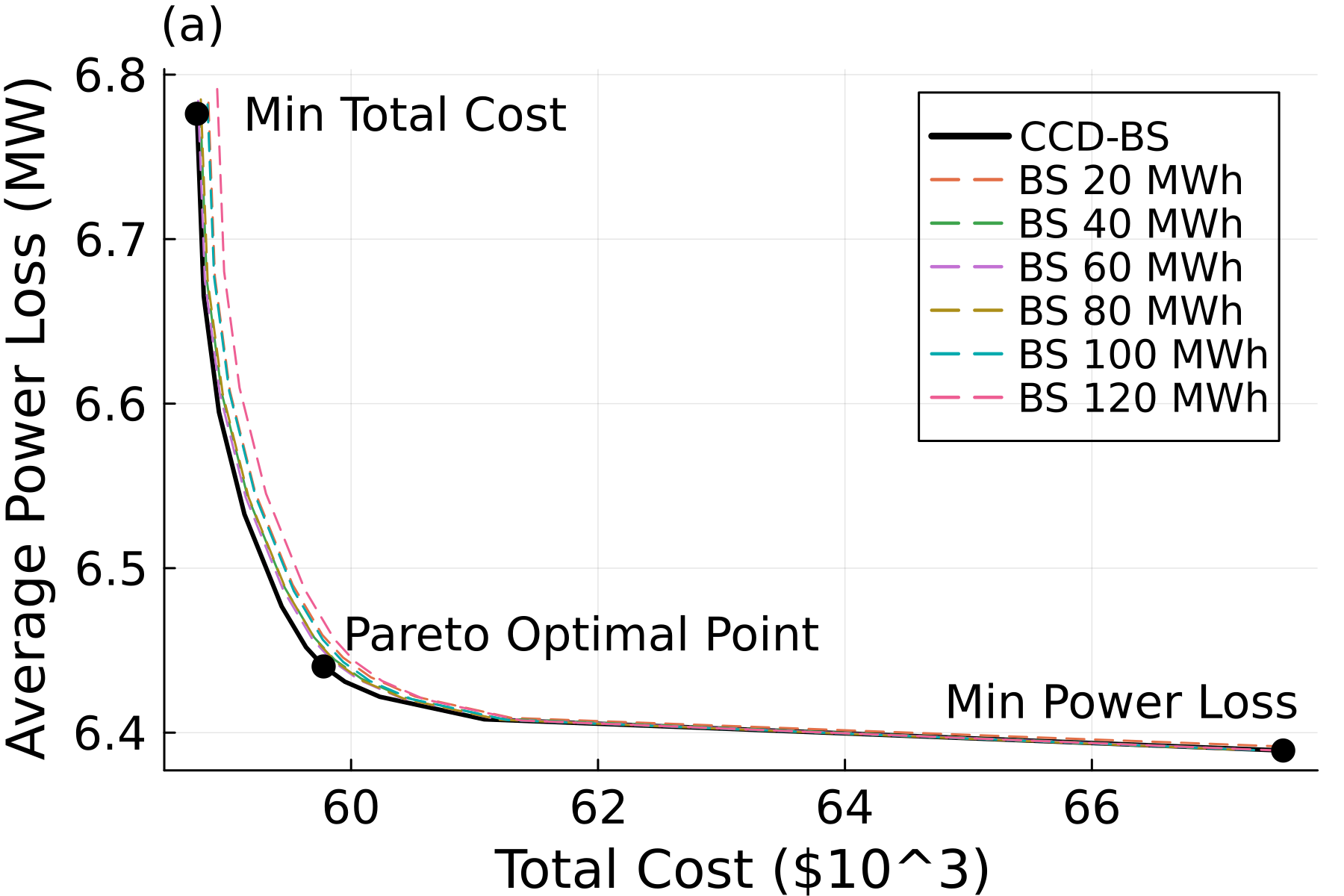}
\includegraphics[width=0.5\linewidth]{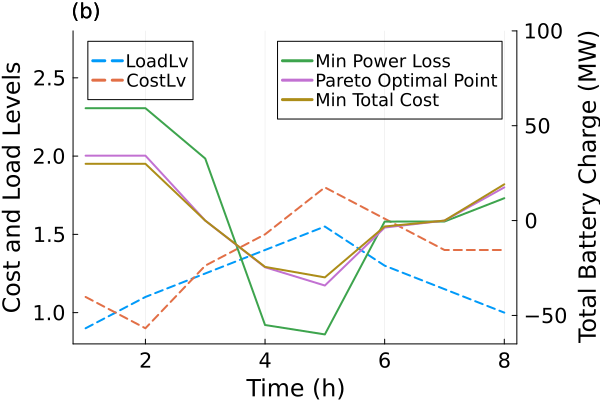}}
\vspace{-0.3cm}
\caption{(a) Pareto front of two objective functions with fixed battery sizes and CCD. \WW{(b) Battery charging/discharging operations in the three selected solutions on the Pareto front.}}
\label{fig_MOjbPF}
\end{figure}

% In Figure~\ref{fig_MOjbPF}~(a) we show a comparison of the Pareto front identified by setting a fixed battery sizes and the one identified using the CCD approach. The dashed lines, representing Pareto fronts from fixed battery sizes, cross each other at some points. The best battery sizes in one area of the objective function space may be dominated in other areas. While the solid black line is always below the other lines, meaning the solution with control co-design battery size that is non-dominated. It takes the lowest total cost under the same power loss levels or has the smallest loss at the same total cost.
\HS{Figure~\ref{fig_MOjbPF}~(a) compares the Pareto fronts derived from fixed battery sizes with those obtained using the CCD approach. The dashed lines depict the fronts for fixed sizes, intersecting in certain regions, indicating that optimal battery sizes for one objective may be suboptimal for others. In contrast, the solid black line, representing the CCD approach, consistently remains below the dashed lines, signifying a non-dominated solution. This indicates that the CCD approach achieves the lowest total cost for a given level of power loss or the least power loss for a given total cost.}

\HS{We selected three points on the Pareto front identified through CCD (refer to Fig.~\ref{fig_MOjbPF}~(a)). The endpoints of the curve represent optimization for a single objective, whereas the midpoint is the Pareto optimal solution that balances both objectives. Figure~\ref{fig_MOjbPF}~(b) illustrates the battery operation at these three points. When minimizing power loss, battery usage is highest; conversely, it is lowest when minimizing total cost. The Pareto optimal solution operates the battery at a level that compromises between minimizing power loss and total cost.}

% \WW{We select three points on the CCD identified Pareto front (Fig.~\ref{fig_MOjbPF}~(a)), where the two end points on the curve aims to optimize only a  single objective, while the middle point is the Pareto optimal solution, given a trade-off between the two objective.Figure~\ref{fig_MOjbPF}~(b) compares the battery operation for these three points.It can be seen that when only power loss is minimized, the batteries are operated the most, while when the total cost is minimized, the batteries are operated least. However, at the Pareto optimal solution aim to operate the battery with a trade-off between the two objectives}

\begin{figure}[htbp]
\centering
\centerline{\includegraphics[scale=0.20,width=0.9\linewidth]{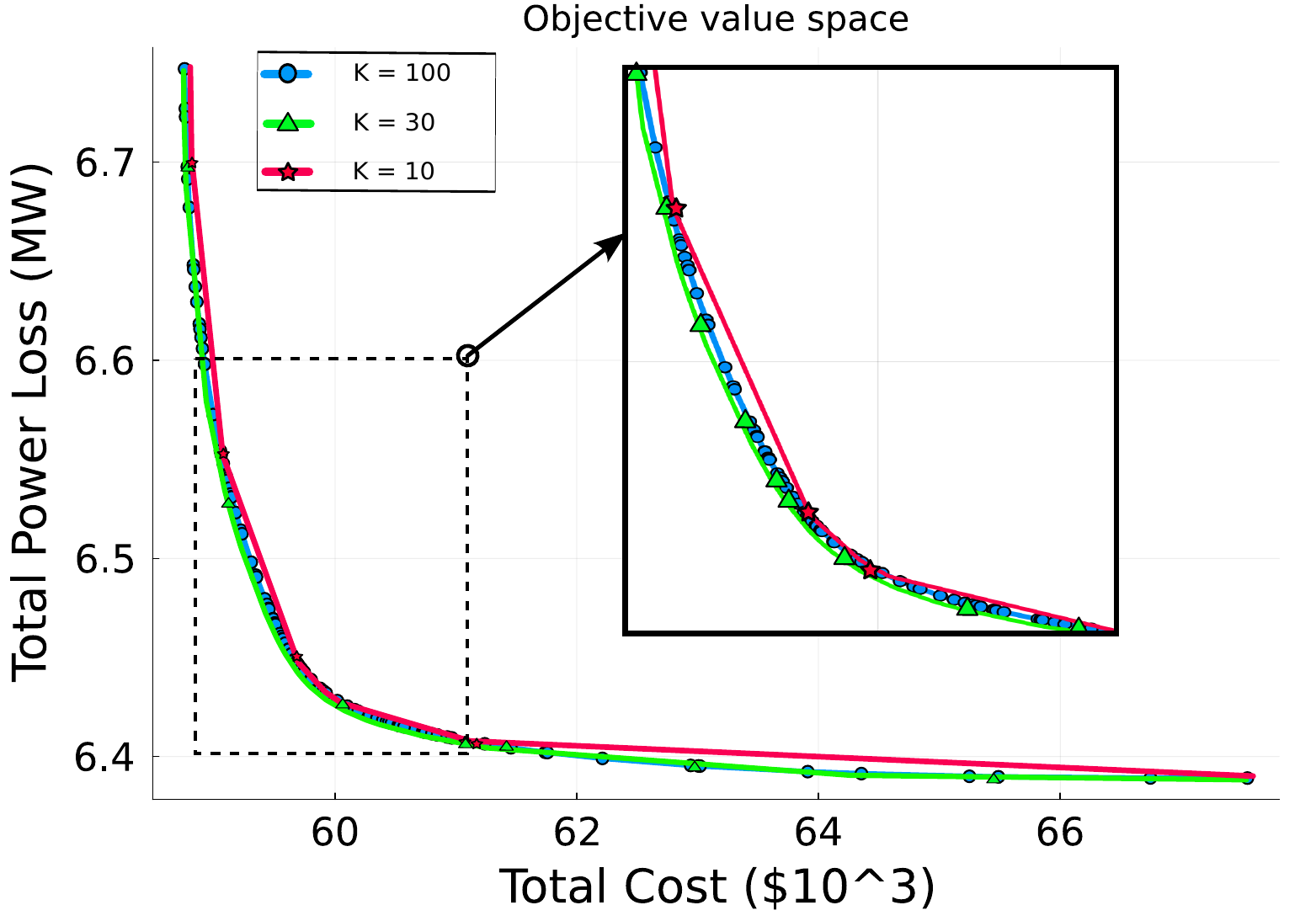}}
% \subfloat[ $K=10$]{
% \includegraphics[width=0.64\linewidth]{figs/f1_f2_K=10_9}
% }

% \subfloat[ $K=30$]{
% \includegraphics[width=0.64\linewidth]{figs/f1_f2_K=30_29}
% }

% \subfloat[ $K=100$]{
% \includegraphics[width=0.64\linewidth]{figs/f1_f2_K=100_99}
% }

\caption{Pareto front of two objective functions with different iterations $K$.}\label{fig:MOO_K1_3}

\end{figure}

Algorithm~\ref{alg:moo} is applied with different number of iterations K=10, 30, 100,to identify weight combination for the Pareto front, shown in Fig.~\ref{fig:MOO_K1_3}. The zoomed view shows the algorithm sampling in the region of maximum change.

\HS{State-of-the-art multi-objective optimization methods like evolutionary and swarm-based algorithms are effective but become costly with larger problems. The graph-based optimization outperforms these methods, offering faster computation and fewer iterations. Applied to the IEEE-9-bus system, it computes the Pareto front with impressive efficiency: for a resolution of K=100, it takes only 17 minutes, and for K=10, just 79.9 seconds, significantly quicker than traditional solvers like IPOPT, which take over 150 seconds for the same task.}

% The state-of-the-art (SOA) approaches for solving multi-objective optimization problem, including evolutionary algorithms, swarm-based algorithms, or intelligent optimization algorithms, can handle complex multi-objective problems but may become computationally expensive and time-consuming as the problem size increases. Compared with SOA approaches, the graph-based formulation in \textit{Plasmo.jl} improves the computational time and reduces the number of iterations significantly. Considering the WSCC-9-bus system with a graph of 224 nodes, 355 edges, and 544 link constraints, the graph-based approach can solve the entire Pareto front within limited iterations. For K=100, the total computational time can be up to 17 minutes, while for K=10, the total computational time is reduced to 79.9 seconds, while it takes more than 150 seconds to solve for the same resolution of Pareto front using the existing solver like \textit{IPOPT}.

\section{CONCLUSIONS}\label{sec:conc}
\HS{Our study introduces a graph-based, multi-objective CCD method for optimizing storage sizing in an AC-grid linked to OWFs via MTDC, considering power generation and transmission controls. Results indicate energy storage can shift demand to off-peak hours, reducing costs, with larger storage preferred for higher demand. A gradient-based framework was used to identify the Pareto Front for minimizing cost and power loss, revealing variable optimal energy storage solutions. Future work will apply this CCD to larger systems and incorporate offshore wind variability, demonstrating the advantages of graph-based optimization.}
\bibliographystyle{IEEEtran}
\bibliography{references}

\begin{thebibliography}{10}
\providecommand{\url}[1]{#1}
\csname url@samestyle\endcsname
\providecommand{\newblock}{\relax}
\providecommand{\bibinfo}[2]{#2}
\providecommand{\BIBentrySTDinterwordspacing}{\spaceskip=0pt\relax}
\providecommand{\BIBentryALTinterwordstretchfactor}{4}
\providecommand{\BIBentryALTinterwordspacing}{\spaceskip=\fontdimen2\font plus
\BIBentryALTinterwordstretchfactor\fontdimen3\font minus \fontdimen4\font\relax}
\providecommand{\BIBforeignlanguage}[2]{{%
\expandafter\ifx\csname l@#1\endcsname\relax
\typeout{** WARNING: IEEEtran.bst: No hyphenation pattern has been}%
\typeout{** loaded for the language `#1'. Using the pattern for}%
\typeout{** the default language instead.}%
\else
\language=\csname l@#1\endcsname
\fi
#2}}
\providecommand{\BIBdecl}{\relax}
\BIBdecl

\bibitem{OffshoreWindReport23}
\BIBentryALTinterwordspacing
W.~Musial, P.~Spitsen, P.~Beiter, P.~Duffy, M.~Shields, D.~Hernando, R.~Hammond, M.~Marquis, J.~King, and S.~Sriharan, ``Offshore wind market report: 2023 edition,'' 2023. [Online]. Available: \url{https://www.energy.gov/eere/wind/articles/offshore-wind-market-report-2023-edition}
\BIBentrySTDinterwordspacing

\bibitem{ansari2020mmc}
J.~A. Ansari, C.~Liu, and S.~A. Khan, ``Mmc based mtdc grids: A detailed review on issues and challenges for operation, control and protection schemes,'' \emph{IEEE Access}, vol.~8, pp. 168\,154--168\,165, 2020.

\bibitem{simao2017challenge}
H.~Sim{\~a}o, W.~Powell, C.~Archer, and W.~Kempton, ``The challenge of integrating offshore wind power in the us electric grid. part ii: Simulation of electricity market operations,'' \emph{Renewable energy}, vol. 103, pp. 418--431, 2017.

\bibitem{zhao2023grid}
C.~Zhao, P.~B. Andersen, C.~Tr{\ae}holt, and S.~Hashemi, ``Grid-connected battery energy storage system: a review on application and integration,'' \emph{Renewable and Sustainable Energy Reviews}, vol. 182, p. 113400, 2023.

\bibitem{Zhao2015ReviewOE}
\BIBentryALTinterwordspacing
H.~Zhao, Q.~Wu, S.~J. Hu, H.~hua Xu, and C.~N. Rasmussen, ``Review of energy storage system for wind power integration support,'' \emph{Applied Energy}, vol. 137, pp. 545--553, 2015. [Online]. Available: \url{https://api.semanticscholar.org/CorpusID:10668291}
\BIBentrySTDinterwordspacing

\bibitem{halwany2022optimal}
N.~Halwany, D.~Pagnani, M.~Ledro, O.~E. Idehe, M.~Marinelli, and L.~Kocewiak, ``Optimal sizing of battery energy storage to enable offshore wind farm black start operation,'' in \emph{21st Wind \& Solar Integration Workshop (WIW 2022)}, vol. 2022.\hskip 1em plus 0.5em minus 0.4em\relax IET, 2022, pp. 232--240.

\bibitem{paul2019multi}
S.~Paul, A.~P. Nath, and Z.~H. Rather, ``A multi-objective planning framework for coordinated generation from offshore wind farm and battery energy storage system,'' \emph{IEEE Transactions on Sustainable Energy}, vol.~11, no.~4, pp. 2087--2097, 2019.

\bibitem{moghaddam2017predictive}
I.~N. Moghaddam, B.~H. Chowdhury, and S.~Mohajeryami, ``Predictive operation and optimal sizing of battery energy storage with high wind energy penetration,'' \emph{IEEE Transactions on Industrial Electronics}, vol.~65, no.~8, pp. 6686--6695, 2017.

\bibitem{allison2014}
J.~T. Allison, T.~Guo, and Z.~Han, ``Co-design of an active suspension using simultaneous dynamic optimization,'' \emph{ASME Journal of Mechanical Design}, vol. 136, no.~8, p. 081003, 2014.

\bibitem{deshmukh2016}
A.~P. "Deshmukh and J.~T. Allison, ``“multidisciplinary dynamic optimization of horizontal axis wind turbine design”.'' \emph{Structural and Multidisciplinary Optimization}, vol.~53, no.~1, p. 15–27, 2016.

\bibitem{yan2009}
H.~S. "Yan and G.~J. Yan, ``"integrated control and mechanism design for the variable input-speed servo four-bar linkages”,'' \emph{"Mechatronics"}, vol.~19, no.~2, p. 274–285, 2009.

\bibitem{liu2020decentralized}
T.~Liu, S.~Azarm, and N.~Chopra, ``Decentralized multisubsystem co-design optimization using direct collocation and decomposition-based methods,'' \emph{Journal of Mechanical Design}, vol. 142, no.~9, 2020.

\bibitem{bhattacharya2021}
A.~Bhattacharya, S.~Vasisht, V.~Adetola, S.~Huang, H.~Sharma, and D.~L. Vrabie, ``Control co-design of commercial building chiller plant using bayesian optimization.'' \emph{Energy and Buildings}, vol. 246, p. 111077, 2021.

\bibitem{nash2021hierarchical}
A.~L. Nash and N.~Jain, ``Hierarchical control co-design using a model fidelity-based decomposition framework,'' \emph{Journal of Mechanical Design}, vol. 143, no.~1, 2021.

\bibitem{nash2021robust}
A.~L. Nash, H.~C. Pangborn, and N.~Jain, ``Robust control co-design with receding-horizon mpc,'' in \emph{2021 American Control Conference (ACC)}.\hskip 1em plus 0.5em minus 0.4em\relax IEEE, 2021, pp. 373--379.

\bibitem{garciasanz2019}
M.~Garcia-Sanz, ``Control co-design: An engineering game changer,'' \emph{Advanced Control for Applications}, vol.~1, no.~1, 2019.

\bibitem{azad2023overview}
S.~Azad and D.~R. Herber, ``An overview of uncertain control co-design formulations,'' \emph{Journal of Mechanical Design}, vol. 145, no.~9, p. 091709, 2023.

\bibitem{allison2024open}
J.~T. Allison, D.~S. Zalkind, and D.~R. Herber, ``Open-loop control co-design of semisubmersible floating offshore wind turbines using linear parameter-varying models,'' \emph{Journal of Mechanical Design}, vol. 146, pp. 041\,704--1, 2024.

\bibitem{sundarrajan2021towards}
A.~K. Sundarrajan and D.~R. Herber, ``Towards a fair comparison between the nested and simultaneous control co-design methods using an active suspension case study,'' in \emph{2021 American Control Conference (ACC)}.\hskip 1em plus 0.5em minus 0.4em\relax IEEE, 2021, pp. 358--365.

\bibitem{allison2014co}
J.~T. Allison, T.~Guo, and Z.~Han, ``Co-design of an active suspension using simultaneous dynamic optimization,'' \emph{Journal of Mechanical Design}, vol. 136, no.~8, p. 081003, 2014.

\bibitem{naik2023pareto}
K.~Naik, S.~Beknalkar, J.~Reed, A.~Mazzoleni, H.~Fathy, and C.~Vermillion, ``Pareto optimal and dual-objective geometric and structural design of an underwater kite for closed-loop flight performance,'' \emph{Journal of Dynamic Systems, Measurement, and Control}, vol. 145, no.~1, p. 011005, 2023.

\bibitem{hallac2017snapvx}
D.~Hallac, C.~Wong, S.~Diamond, A.~Sharang, R.~Sosic, S.~Boyd, and J.~Leskovec, ``Snapvx: A network-based convex optimization solver,'' \emph{The Journal of Machine Learning Research}, vol.~18, no.~1, pp. 110--114, 2017.

\bibitem{jalving2017graph}
J.~Jalving, S.~Abhyankar, K.~Kim, M.~Hereld, and V.~M. Zavala, ``A graph-based computational framework for simulation and optimisation of coupled infrastructure networks,'' \emph{IET Generation, Transmission \& Distribution}, vol.~11, no.~12, pp. 3163--3176, 2017.

\bibitem{berger2021remote}
M.~Berger, D.~Radu, G.~Detienne, T.~Deschuyteneer, A.~Richel, and D.~Ernst, ``Remote renewable hubs for carbon-neutral synthetic fuel production,'' \emph{Frontiers in Energy Research}, p. 200, 2021.

\bibitem{abhyankar2018petsc}
S.~Abhyankar, J.~Brown, E.~M. Constantinescu, D.~Ghosh, B.~F. Smith, and H.~Zhang, ``Petsc/ts: A modern scalable ode/dae solver library,'' \emph{arXiv preprint arXiv:1806.01437}, 2018.

\bibitem{jalving2019graph}
J.~Jalving, Y.~Cao, and V.~M. Zavala, ``Graph-based modeling and simulation of complex systems,'' \emph{Computers \& Chemical Engineering}, vol. 125, pp. 134--154, 2019.

\bibitem{jalving2022graph}
J.~Jalving, S.~Shin, and V.~M. Zavala, ``A graph-based modeling abstraction for optimization: Concepts and implementation in plasmo. jl,'' \emph{Mathematical Programming Computation}, vol.~14, no.~4, pp. 699--747, 2022.

\bibitem{gu2022min}
A.~Gu, S.~Lu, P.~Ram, and L.~Weng, ``Min-max bilevel multi-objective optimization with applications in machine learning,'' \emph{arXiv preprint arXiv:2203.01924}, 2022.

\bibitem{bresesti2007hvdc}
P.~Bresesti, W.~L. Kling, R.~L. Hendriks, and R.~Vailati, ``Hvdc connection of offshore wind farms to the transmission system,'' \emph{IEEE Transactions on energy conversion}, vol.~22, no.~1, pp. 37--43, 2007.

\bibitem{liang2011operation}
J.~Liang, T.~Jing, O.~Gomis-Bellmunt, J.~Ekanayake, and N.~Jenkins, ``Operation and control of multiterminal hvdc transmission for offshore wind farms,'' \emph{IEEE Transactions on Power Delivery}, vol.~26, no.~4, pp. 2596--2604, 2011.

\bibitem{baradar2012power}
M.~Baradar, M.~Ghandhari, D.~Van~Hertem, and A.~Kargarian, ``Power flow calculation of hybrid ac/dc power systems,'' in \emph{2012 IEEE Power and Energy Society General Meeting}.\hskip 1em plus 0.5em minus 0.4em\relax IEEE, 2012, pp. 1--6.

\bibitem{zhou2019optimal}
Y.~Zhou, L.~Zhao, W.-J. Lee, Z.~Zhang, and P.~Wang, ``Optimal power flow in hybrid ac and multi-terminal hvdc networks with offshore wind farm integration based on semidefinite programming,'' in \emph{2019 IEEE Innovative Smart Grid Technologies-Asia (ISGT Asia)}.\hskip 1em plus 0.5em minus 0.4em\relax IEEE, 2019, pp. 207--212.

\bibitem{jabr2006radial}
R.~A. Jabr, ``Radial distribution load flow using conic programming,'' \emph{IEEE transactions on power systems}, vol.~21, no.~3, pp. 1458--1459, 2006.

\bibitem{ergun2019optimal}
H.~Ergun, J.~Dave, D.~Van~Hertem, and F.~Geth, ``Optimal power flow for ac--dc grids: Formulation, convex relaxation, linear approximation, and implementation,'' \emph{IEEE transactions on power systems}, vol.~34, no.~4, pp. 2980--2990, 2019.

\end{thebibliography}

\end{document}